\documentclass[runningheads]{llncs}

\usepackage{graphicx}

\def\BibTeX{{\rm B\kern-.05em{\sc i\kern-.025em b}\kern-.08emT\kern-.1667em\lower.7ex\hbox{E}\kern-.125emX}}

\usepackage{balance}       
\usepackage{graphics}      
\usepackage[T1]{fontenc}   
\usepackage{txfonts}
\usepackage{mathptmx}
\usepackage{color}
\usepackage{textcomp}
\usepackage{multirow}
\usepackage{array}

\usepackage[algoruled,vlined,linesnumbered]{algorithm2e}

\usepackage{ccicons}          

\usepackage{todonotes}

\usepackage{comment}

\def\norad{NORA-D}
\def\norat{NORA-T}

\DeclareMathDelimiter{(}{\mathopen} {operators}{"28}{largesymbols}{"00}
\DeclareMathDelimiter{)}{\mathclose} {operators}{"29}{largesymbols}{"01}
\DeclareMathDelimiter{]}{\mathopen} {operators}{"5D}{largesymbols}{"00}
\DeclareMathDelimiter{[}{\mathclose} {operators}{"5B}{largesymbols}{"01}
\DeclareMathDelimiter{=}{\mathrel} {operators}{"3D}{largesymbols}{"00}
\DeclareMathDelimiter{+}{\mathrel} {operators}{"2B}{largesymbols}{"01}

\begin{document}

\title{Client Network: An Interactive Model for Predicting New Clients}
%
%
\author{Massimiliano Mattetti\inst{1} \and
Akihiro Kishimoto\inst{1} \and
Adi Botea\inst{1} \and
Elizabeth Daly\inst{1} \and
Inge Vejsbjerg\inst{1} \and
Bei Chen\inst{1} \and
\"{O}znur Alkan\inst{1}}
\authorrunning{M. Mattetti et al.}
%
\institute{IBM Research - Ireland\\
\email{massimiliano.mattetti@ibm.com \and \{akihirok,adibotea,elizabeth.daly,ingevejs,oalkan2,beichen2\}@ie.ibm.com}}

\newcommand{\towrite}[1]{\hidecomments{\vspace{1em}\todo[inline,color=black!20!white]{\textbf{TODO:} #1}\vspace{1em}}}

\newcommand{\hidecomments}[1]{#1}

\newcommand{\ab}[1]{\hidecomments{\vspace{1em}\todo[inline,color=blue!20!white]{\textbf{Adi:} #1}\vspace{1em}}}
\newcommand{\ed}[1]{\hidecomments{\vspace{1em}\todo[inline,color=red!20!white]{\textbf{Elizabeth:} #1}\vspace{1em}}}
\newcommand{\oa}[1]{\hidecomments{\vspace{1em}\todo[inline,color=green!20!white]{\textbf{Oznur:} #1}\vspace{1em}}}
\newcommand{\ak}[1]{\hidecomments{\vspace{1em}\todo[inline,color=yellow!20!white]{\textbf{Kishi:} #1}\vspace{1em}}}
\newcommand{\iv}[1]{\hidecomments{\vspace{1em}\todo[inline,color=orange!20!white]{\textbf{Inge:} #1}\vspace{1em}}}
\newcommand{\bc}[1]{\hidecomments{\vspace{1em}\todo[inline,color=pink!20!white]{\textbf{Bei:} #1}\vspace{1em}}}
\newcommand{\mm}[1]{\hidecomments{\vspace{1em}\todo[inline,color=orange!20!white]{\textbf{Massimiliano:} #1}\vspace{1em}}}

\newcommand{\optarg}[1]{\ifthenelse{\isempty{#1}}{}{(#1)}}

\newcommand{\hide}[1]{}

\maketitle

\begin{abstract}
	Understanding prospective clients becomes increasingly important as companies
	aim to enlarge their market bases. Traditional approaches typically treat each
	client in isolation, either studying its interactions or similarities with
	existing clients. We propose the Client Network, which considers the entire
	client ecosystem to predict the success of sale pitches for targeted clients
	by complex network analysis. It combines a novel ranking algorithm with data
	visualization and navigation. Based on historical interaction data between
	companies and clients, the Client Network leverages organizational
	connectivity to locate the optimal paths to prospective clients. The user
	interface supports exploring the client ecosystem and performing
	sales-essential tasks. Our experiments and user interviews demonstrate the
	effectiveness of the Client Network and its success in supporting sellers'
	day-to-day tasks.

\end{abstract}

\section{Introduction}

Identifying new potential clients is an important and time-consuming task for
many organizations. Sellers have to prospect for new clients and figure out
which companies might be more interested in becoming a client. Sellers may also
have to follow up on marketing responses where potential clients
may have previously expressed an interest in a product by downloading a white
paper or asking for additional information as a follow up to a marketing
campaign. With limited time and large numbers of follow-ups to pursue, sellers
may need assistance in prioritizing which leads represent likely conversions.
Typical approaches to client prospecting use the notion of similarity to
identify clients who have shared attributes to the organization's existing
clients. Prioritization methods tend to filter out low likelihood interactions
or identify spam-like behavior to focus on interactions that express the most
interest. These solutions evaluate the organization or interactions in isolation
from the existing client relationships. Knowledge and trust of a brand is
something that is built up over time with clients and can play an important role
in a client's decision to purchase from a company in the future. This influence
does not need to be limited to the existing clients, given that the business
world is a small world where key professionals move between organizations. When
a person takes on a new role in an organization they do not just start afresh,
they also bring with them their knowledge, experiences and relationships. This
can tell us something about the companies that they are moving between: it could
be that either they are in a similar industry space or have a similar corporate
culture. We can harness the fact that they may bring with them knowledge and
experience of products and services used in their prior organization. If we can
find these companies, they can represent valuable prospective clients for an
organization to approach.

In this paper we explore the research question of how to leverage latent
information in the relationship graph to predict prospective clients
and also surface the relationships to sellers, to provide tangible
explanations and give context for the predictions. We present our solution,
Client Network, which uses organizational connectivity to build a
network with different types of links with different meanings and different
weights. It combines a novel ranking algorithm with data visualization to allow
the sellers to understand the client ecosystem better. We tuned the algorithm by
analyzing our connectivity with existing clients assuming the client link is
missing in order to be able to use this ranking to predict the utility of the
prospective client.

Our contributions in this paper are as follows: A scalable network algorithm
that supports heterogenous network relationships to rank prospective clients in
order to allow sellers to focus on leads with the highest probability of
success; and an interactive visualisation tool which supports exploration and provides context
supporting the predictions allowing them to interpret the ranking and better
inform the seller.

\section{Background} \label{sec:background}
Given a network of companies connected together via multiple relationships, the
problem of recommending prospective clients can be formulated as predicting
whether the link \emph{client} appears in the network or not. Such problem is a
variant of the standard \emph{link prediction problem} which aims to predict the
formation of any type of relationships in a network. It has extensively been
studied in various fields, such as web
science~\cite{Adafre:2005:DML:1134271.1134284,Zhu:2002:UMM:513338.513381},
healthcare and
biology~\cite{Johnson:2012:AML:2450057.2450061,DBLP:journals/netmahib/AlmansooriGJEMJAR12,airoldi06,DBLP:conf/lion/Freschi09},
and recommender
systems~\cite{Aiello:2012:FPH:2180861.2180866,MORI201210402,Wu:2013:PPR:2433396.2433404,ASI:ASI20591,Huang:2005:LPA:1065385.1065415}.

Formally, the link prediction task can be formulated as
follows~\cite{ASI:ASI20591}. Given a network $G(V,E)$, let edge $e = (u, v) \in
E$ represent an interaction between nodes $u, v \in V$ at time $t(e)$. The
multiple interactions between $u$ and $v$ are recorded as parallel edges. For $t
\leq t'$, let $G\lbrack t, t'\rbrack$ denote the subgraph of $G$ restricted to
the edges between $t$ and $t'$. For \emph{training interval} $\lbrack t_0, t_0'
\rbrack$, the link prediction task is to predict a list of edges occur in the
network $G \lbrack t_1, t_1' \rbrack$, where $t_1 > t_0'$. 

Various techniques exist for link prediction which differs in model complexity,
prediction performance, scalability, and generalization ability (see
~\cite{DBLP:journals/corr/WangXWZ14,Martinez:2016:SLP:3022634.3012704,DBLP:books/sp/social2011/HasanZ11}
for an overview). Topology-based link prediction
methods~\cite{DBLP:journals/corr/WangXWZ14} are unsupervised approaches that
leverage the \emph{latent information} contained in the network topology to
assign a score to each pair of nodes. Algorithms in this category are generally
divided into $local$ and $global$ similarity-based
approaches~\cite{Martinez:2016:SLP:3022634.3012704}. Local approaches, such as
Common Neighbors, the Jaccard Coefficient and the Adamic-Adar
Coefficient~\cite{AdaAda03}, use node neighbourhood-related structural
information to compute similarity among nodes. They are simple and fast to
compute ~\cite{Martinez:2016:SLP:3022634.3012704}. However, their performance in
non-small-networks~\cite{ASI:ASI20591}, where links can form between nodes at
distances greater than two, is poor. Global approaches, on the other hand, use
the whole network topological information to score each link. They have strong
predictive power but at higher computational cost. Path-based algorithms, such
as the Katz index~\cite{RePEc:spr:psycho:v:18:y:1953:i:1:p:39-43}, and
algorithms based on random walks, such as PageRank~\cite{page1999},
SimRank~\cite{Jeh:2002:SMS:775047.775126}, Hitting Time~\cite{ASI:ASI20591} and
PropFlow~\cite{Lichtenwalter:2010:NPM:1835804.1835837}, fall under the umbrella
of global topology-based approaches.

A wide range of real-world systems can be modelled as \emph{heterogeneous
information networks} (HIN), such as human social activities, communication and
computer systems, and biological networks. Formally a HIN is defined as a
network of multiple types of nodes (e.g., authors, conferences and papers) and
multiple types of edges (e.g., co-author, author-write-paper and
paper-published-in-conference)~\cite{Shi:2017:SHI:3024189.3024193}.

Initial approaches to the link prediction problem in HIN utilized the same
algorithms used in the homogeneous ones, with no changes. These algorithms were
not designed to take into account the dependency patterns across types that
exist in heterogeneous networks. Such approaches treat all relationships equally
or separately study homogeneous projections of the networks, completely ignoring
information about the different topology or the different formation mechanism
that each type of edge may have. Later approaches introduced new extensions to
classical link prediction
algorithms~\cite{Davis:2011:MLP:2055438.2055676,Yang:2012:PLM:2471881.2472620}
which improved their performance in HIN. More recently, a meta-structure known
as a \emph{meta path}~\cite{journals/pvldb/SunHYYW11} has been proposed in order
to account for the semantic of the different types of relationships.
Furthermore, a new category of algorithms has arisen which leverages the concept
of meta path~\cite{journals/kais/ShiLYW16,Shakibian:2017:SR}.

\begin{figure}[h]
	\centering
	\includegraphics[width=0.8\columnwidth]{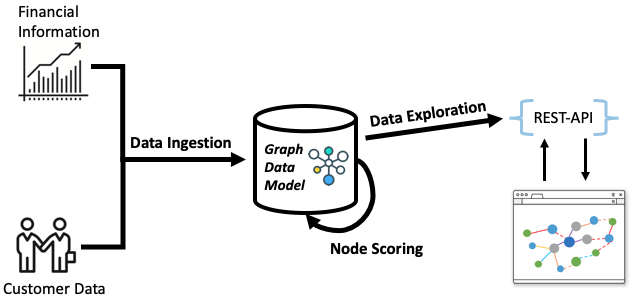}
	\caption{Application Workflow}~\label{fig:solution-workflow}
\end{figure}

\section{The Proposed Solution: Client Network} \label{sec:solution}

Our solution, Client Network, aims at supporting sellers to identify prospective
clients. It leverages past interactions that took place within an ecosystem of
people and organizations to build a network connecting the seller's company to
any other company in the ecosystem. It employs a novel ranking algorithm for
computing the likelihood of turning a company into a new client and enables the
seller to explore the network with an interactive UI. Hereafter we refer to the
seller's company with the term \emph{root}.

{\bf Data Model.} The Client Network can work with any type of network providing
it has the following characteristics: i) a subset of nodes represent companies;
ii) one of the nodes of type company is marked as \emph{root}; iii) a
relationship of type \emph{client} exists between the root node and other nodes
of type company.
%

Thus, the Client Network is a generic solution, with no additional assumptions
about the types of nodes and the types of relations, for example.

However, for clarity, we use details from an actual network in the presentation. 
This is also used as input to our system in the evaluation. It is a network that
combines information about the clients of our organization with financial
information about banks, corporations, investment managers and more. Being 
confidential data, we are not able to publish the dataset used in the evaluation. 
The network contains two types of nodes, company and person. The latter is 
for professionals with decisional power in their companies. Each job role of a professional,
either current or former, is represented as an edge between the corresponding
person node and company node.

Job roles are divided into two types: \emph{board member} and \emph{executive}.
We further attach to job-role edges a label such as \emph{current} or
\emph{former}. For instance, given a company and its current CTO, the
corresponding company and person nodes are connected with an edge labelled as
\emph{current executive}. Likewise, the node of a professional that has served
in the company's board in the past is connected to the company's node through an
edge labelled as \emph{former board member}.

In addition to the connections capturing the organizational people flow, the
network contains \emph{business to business} (B2B) relationships, such as
\emph{sponsor}, \emph{subsidiary} and \emph{investor}, which can be in different
states, such as \emph{pending}, \emph{cancelled} and \emph{prior}.

Despite being a B2B relationship, the \emph{client} relationship differs from
the others because it does not have a state. In other words, a company is either
a current client of the root company or not. Former clients\footnote{In this
work, a former client is a company that has not purchased any product/service
produced/provided by the root company in the last 5 years.} are equivalent to
non-client companies from the seller's point of view. Note that a client
relationship always involves the root, being an edge between the root and the
client at hand. It is worth mentioning that there is a significant imbalance in
the distribution of relationships and nodes. For example, of the 11.5 million
nodes in the network, two-thirds are organizations whereas the remaining third
are professionals. Furthermore, within the organizations there is a ratio of 1
to 14 between the client and non-client nodes, respectively.

Finally, multiple relationships can exist between two entities, that is, the
Client Network allows multiple edges between a pair of nodes.


{\bf Solution Overview.} The Client Network includes two main components: the
\emph{Analyzer} and the \emph{Interactive Visualizer}.

The Analyzer extracts firmographic data and biographical information about
professionals from relational data sources, converts the data into a graph data
model and finally loads it into a database. Once the structure of the network is
ready, our NORA algorithm, described in Section~\ref{subsec:ranking}, assigns a
score to each node in the	network. These scores are then stored as attributes of
the node objects in the database.

The Interactive Visualizer allows users to interact with the network. Possible
interactions include exploring the neighborhood of a node, retrieving the
subgraph connecting a node to the root node, and ranking a list of companies,
given as input, using NORA. These functionalities are backed by a set of REST
APIs which have been designed to support any sales application in exploring the
network data, therefore allowing an easier integration of our model with
existing sellers' tools.

The client ecosystem is a very dynamic environment. A new client acquisition, an
individual who moves to a different company and a company that invests in
another company are all examples of events that contribute to the evolution of
the network. The Client Network requires an up-to-date snapshot of the network
in order to provide a high prediction accuracy. Hence, a full import of the
data, as well as a new scoring of the nodes, have to be performed periodically.
Figure~\ref{fig:solution-workflow} describes the full workflow of our solution,
where data ingestion and node scoring tasks are performed by the Analyzer and
data exploration is enabled with the Interactive Visualizer.


{\bf User interface.}
An API and UI have been developed to support navigating through the complex network of companies and individuals.
Our aim was to facilitate
the exploration of the network context while maintaining a task-oriented focus.
Users can discover prospective clients with connections to the root company,
explore each client's relationships and receive
a measure of the success chances of a sales pitch
to that client.
We envisioned that the Client Network
would help
sellers in several important tasks, such as:
following up on a list of leads from marketing;
exploring the connections of a recently acquired client;
and finding out more about a social media interaction.

Initial discussions with domain experts revealed that one concern for the
root organization was the amount of time that the sellers spend prospecting for
\emph{whitespace}\footnote{Term used internally by sellers to refer to
	prospective clients.} companies using their existing tools like LinkedIn. Using the
Client Network will allow sellers to save time when deciding which marketing
leads they should prioritize, as well as to make discoveries which could help
them shape their approach to the client and increase the likelihood of a
successful sales opportunity outcome. The seller can get a feeling for how the
individual company fits into a larger and connected client ecosystem relating
back to their company.


The user interface is composed of a general landing and information area, a
search interface to find companies, company ranking list views and a view to
explore a company and its connections in more detail. The user interface
consists of a React.js web application which communicates with a Java API that
supplies it with the data to power the search and the visualizations. The UI is
built responsively, so that it will respond to screen size and show appropriate
styling and information depending on the device screen size. The network graph
is visualized using the vis.js library~\cite{Visjs} and it uses a force-directed
layout, with controls for zooming and resetting the graph.

The search functionality allows the user to search for a company name. On searching for a company, the user is
presented with a list of results which shows company features such as the
company name, ranking, probability of success, status, location, and year
founded, allowing the user to disambiguate between similarly named companies.
Once the targeted company is identified, the user can click through to access a
detailed view of that company, or they can add them to a ranking list.

\begin{figure*}
	\begin{tabular}{ccc}
		\fcolorbox{black}{white}{\includegraphics[width=.45\columnwidth]{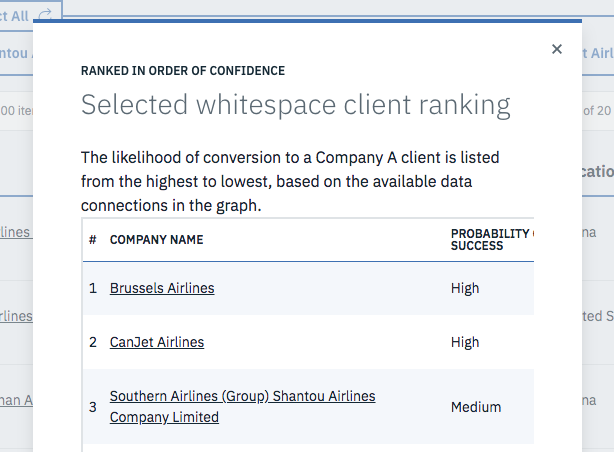}}
		  & &
		\fcolorbox{black}{white}{\includegraphics[width=.5\columnwidth]{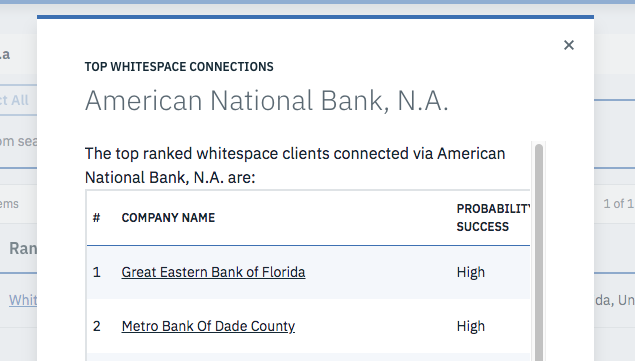}}
	\end{tabular}
	\caption{Left: Company ranking. Right: Top Whitespace Companies}
	\label{fig:rankingAndWhitespace}
\end{figure*}

Sellers receive lists of prospective clients from the marketing team or their
managers. Deciding which of these clients is a likely prospect and should be
approached as a priority can be time-consuming. The seller can input a list of
companies and the Client Network will return the list ranked by the confidence
level of turning a company into a new client. The seller searches for each
company in turn and adds them to the list. The system processes this list,
returning a modal window with the list of companies. The list is ranked, showing
companies assumed to have a higher chance of success of becoming a future client
first, as presented in Figure~\ref{fig:rankingAndWhitespace} left.


Depending on factors such as the product they are selling or the country they
are targeting, some sellers may receive fewer leads than others. In the case of
a seller struggling to find prospective clients, a good starting point would
be to explore the connections of a recently acquired client. An existing client
can be used as a starting-off-point to prospect for whitespace clients that are
connected to it. Thus, the Client Network turns a newly acquired client into an
opportunity for approaching whitespace clients which were not taken into
consideration previously due to a low connectivity to the root company. The
seller searches for the company and can further investigate the ``Whitespace
connections''. A ranked list of whitespace clients that are connected to the
newly acquired client is returned to the seller
(Figure~\ref{fig:rankingAndWhitespace} right) allowing them to quickly identify companies in
that ecosystem that have a high chance of conversion to a client.

\begin{figure*}
	\centering
	\fcolorbox{black}{white}{\includegraphics[width=1\columnwidth]{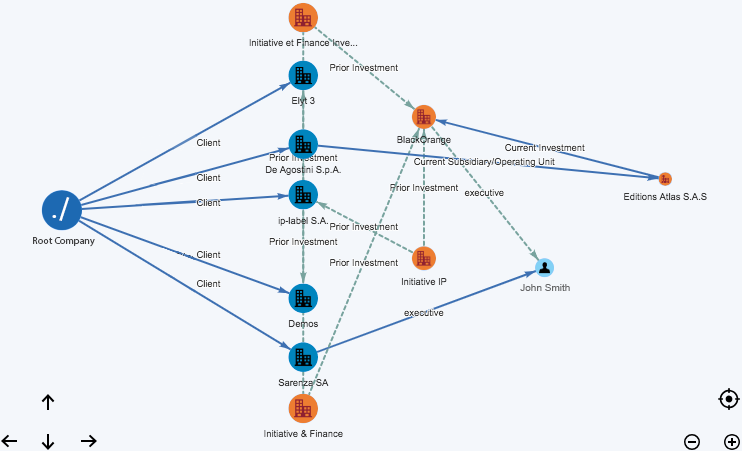}}
	\caption{A detailed view of the company BlackOrange and its connections}
	\label{fig:companydetail1}
\end{figure*}

In the third scenario, the seller can find information about the positioning of
the company within the client ecosystem and use company descriptions and
professional biographies to find out more. The seller can select a company to
see a detailed view of that company and its associated subgraph, which
visualizes how it links back to the root company, as shown in
Figure~\ref{fig:companydetail1}. The seller can further see
details about companies and professionals that are connected by a series of
relationships. Each relationship is labelled with its type. Current
relationships are displayed with a solid line and previous relationships with a
dashed line. The size of the node is dictated by its connectivity with the root
company and is related to the size of the other nodes in the subgraph, that is
to say it is scaled based on the scores of the other nodes in the current
subgraph.

An important feature is the interaction design to allow the
sellers to make sense of a large graph. It has been suggested that producing a
good visualization gets harder as the graph gets larger with difficulties in the
syntactic (e.g., avoiding occlusion), and semantic (e.g., highlighting the
important features of the underlying network) areas~\cite{ware2002cognitive}.
With this in mind, the decision was taken to limit the initial view to a
subgraph so that the user did not suffer from information overload and that
initial view of the graph is not as overwhelming as it would be if the subgraph
was too large for the window in which it is presented. Not every link between
the root company and the target company is visualized, but just a selection of
the shortest paths. The user can explore the graph in any direction that they
find interesting by requesting the visualization of any further connections that
the selected node in the subgraph may have. Information about the target company
and each other node in the subgraph is available below the diagram in the form
of short biographies of professionals and company descriptions.

A good visualization should emphasize the readability and promote the
understanding of the underlying relationships. Our aim with this visualization
is to afford users the opportunity to understand the intrinsic information
contained in the structure of the network, enabling them to better understand
the data that is available about the companies and professionals. The external
image of the graph becomes a kind of \emph{information
	store}~\cite{zhang1994representations}, allowing the sellers to use it as a
space for visual problem solving.

\subsubsection{Ranking Mechanism} \label{subsec:ranking}

Marketing campaigns allow companies to capture the interest of prospective
clients. This interest is converted into leads passed to the sales division. The
seller who is supposed to follow up on these leads has to go through a careful
skimming phase which involves prospecting and prioritizing the companies in the
list. Our domain experts estimate that a seller spends over 30\% of their time
on these activities.

To support the seller in taking a decision about which companies should have the
highest priority and which ones may not be worth pursuing, the Client Network
assigns a score to each company in the network. These scores are computed using
our Node Relevance Algorithm (NORA), which computes a \emph{flow value} of each
node in a graph. Then, nodes can be ranked (ordered decreasingly) based on the
scores. The flow value of a node measures the strength of the relation to the
source node (seller's company). See details about the computation of flow values
later in this section.

Intuitively, the sellers should approach a company that has a close connection to their company. 
Additionally, a company that has multiple connections to the sellers' company should be
considered as a strong candidate. 
NORA's flow values attempt to satisfy these criteria
by considering the subset of the global topological structure of the Client Network. 

Algorithm~\ref{alg:nora} shows the main steps of NORA in pseudocode. At line 1,
method {\tt Analyze\-Network} has a three-fold purpose: compute an ordering of the
nodes $\Delta$, and a filtered set of edges $E' \subseteq E$. Secondly, edges
in $E'$ become directed edges, as explained later in this section.
Finally, $E'$ allows NORA to ensure the convergence of flow values with low computational
complexity. 
We will
introduce two approaches of this step, leading to two variations of our
algorithm, called {\norad} and {\norat}.

\begin{algorithm}[t!] \DontPrintSemicolon
	\small
	\caption{{\tt NORA} ranking technique} \label{alg:nora}
	\SetKwInOut{Input}{input} \SetKwInOut{Output}{output} \Input{graph $G =
	(V,E)$, start node $s$} \Output{Scores of all nodes, to be used for ranking}
	$(\Delta,E') \leftarrow \mbox{AnalyzeNetwork}(G,s)$ \;
        \For {$v \in V$ in the order given by $\Delta$} {$F(v) = \gamma \times \Sigma_{p \in P}
             \frac{F(p)}{\mbox{od}(p)}$, where  $P$ is the set of all parent nodes in $G'=(V,E')$
             and $\gamma < 1$ is a discount factor. \; \label{pseudocode:flow-val-form}}        
	\Return $F$
\end{algorithm}

The next step of NORA (lines 2--3 in Algorithm~\ref{alg:nora}) is to compute a
function $F: V \rightarrow \Re_+$, called the flow-value function. A lookup
table is a simple and easy implementation for $F$.
Nodes are parsed in the order given by $\Delta$. It remains to describe the
actual formula to compute the flow value of a given node $v$
(line~\ref{pseudocode:flow-val-form} in the pseudocode). Given a node $p$, let
$\mbox{od}(p)$ be the number of outgoing directed edges from $p$, in graph $G'$.
For a given node $v$, let $P$ be the set of all parent nodes in $G'$. The flow
of $v$ is defined as: $F(v) = \gamma \times \Sigma_{p \in P}
\frac{F(p)}{\mbox{od}(p)},$ where $\gamma < 1$ is a discount factor.
Intuitively, a parent node $p$ will distribute its flow value among its children
in $G'$. In the formula above, the value is shared equally among the children
nodes of $p$. A child node accumulates flow values from all its parent nodes in
$G'$. The discount factor $\gamma$ implements the intuition that, if a node $v$
is further away from $s$, then its flow value should be smaller.

Next we present {\norad} and {\norat}, our algorithmic versions that differ in
the way they implement step 1 in Algorithm~\ref{alg:nora}.

{\bf \norad.} In this variant of our algorithm, step 1, corresponding to {\tt
Analyze\-Network}, performs a one-to-many shortest-path search to produce its
output. We discuss two possible implementations of shortest-path search, one
based on Dijkstra's algorithm and one based on breadth-first search.

With no assumptions made about whether the edges have a uniform cost or not, the
first implementation runs Dijkstra's
algorithm~\cite{Dijkstra:1959:NTP:2722880.2722945} from the source node $s$. The
Dijkstra algorithm takes as input a graph, such as $G' = (V,E')$, and a node $s
\in V$. It computes the distance (i.e., the cost of a minimum-cost path) from
$v$ to any other node in the graph. As such, Dijkstra's algorithm is a
one-to-many distance computation technique.

\begin{algorithm}[t!] \DontPrintSemicolon
	\small
	\caption{{\tt Extended Dijkstra}} \label{alg:ext-dijkstra}
	\SetKwInOut{Input}{input} \SetKwInOut{Output}{output} \Input{graph $G =
	(V,E)$, start node $s$} \Output{Ordering of nodes $\Delta$, in increasing
	order of the distance from $s$; Edges that belong to shortest paths from
	$s$, in the data structure $E'$} $g(s) \leftarrow 0$ \; \For {$v \in V, v
	\neq s$} {$g(v) \leftarrow \infty$ \;} $P \leftarrow \{s\}$ \;
	\While {$P \neq \emptyset$} {$n \leftarrow \mbox{pop}(P)$ \; append $n$ to
		$\Delta$ \;
		\For {$(n,c) \in E$} {\If {$g(c) > g(n) + \mbox{cost}(n,c)$} {$g(c)
			\leftarrow g(n) + \mbox{cost}(n,c)$ \; insert $c$ into $P$, unless
			this has been done before \; $\mbox{MarkedIncomingEdges}(c)
			\leftarrow \emptyset$ \;} $(n,c) \rightarrow
			\mbox{MarkedIncomingEdges}(c)$ \;}} $E' \leftarrow \cup_{v \in V, v
			\neq s} \mbox{MarkedIncomingEdges}(v)$ \; \Return $\Delta,E'$
\end{algorithm}

A standard implementation of Dijkstra's algorithm returns, for each node in $v
	\in V$, the cost of an optimal path from $s$ to $v$. We have slightly
	modified the Dijkstra implementation to return additional information beside
	the optimal costs. Specifically, we mark all edges in $E$ with the property
	that they belong to an optimal path from $s$ to some node $v$. More
	formally, let $\mbox{Opt}(a,b)$ be the set of all optimal paths from a node
	$a$ to a node $b$. An edge $e$ is marked iff $\exists v \in V, \exists \pi
	\in \mbox{Opt}(s,v) \mbox{ such that } e \in \pi$. Marked edges are added
	into the subset $E'$. We further assign a direction to the edges in $E'$,
	from the node with a smaller cost to the node with the larger cost. As such,
	the graph $G' = (V,E')$ is a directed acyclic graph (DAG).

Algorithm~\ref{alg:ext-dijkstra} shows the Dijkstra's algorithm in pseudocode,
together with our extension that marks all edges that we want to keep after
filtering. Dijkstra's algorithm uses a priority list $P$ populated with elements
$n \in V$, ordered on the cost $g(n)$. The cost $g(n)$, also called the $g$
value of the node $n$, is the cost from the starting node to the current node
$n$. Popping a node from $P$ returns and removes from $P$ an element with a
smallest $g$ value. The algorithm expands each node once, in an increasing order
of the $g$ value. Expanding a node $n$ updates the $g$ value of each successor
$c$, if reaching $c$ through $n$ is a shortest path from $s$ to $c$ among all
paths from $s$ to $c$ explored so far. At an expansion step, we also insert
successor nodes into the priority list, unless a given successor node has been
inserted into $P$ at a previous time. Our extensions, which keep track of
$\Delta$ and $E'$, are shown at lines 7, 12, 13, 14 and 15.

On a uniform-cost graph, we can replace Dijkstra's algorithm with breadth-first
search. This reduces the complexity of method {\tt Analyze\-Network}, as
discussed later in this section.

\begin{algorithm}[t!] \DontPrintSemicolon
	\small
	\caption{{\tt AnalyzeNetwork (\norad)}} \label{alg:filter-edges}
	\SetKwInOut{Input}{input} \SetKwInOut{Output}{output} \Input{graph $G =
	(V,E)$, start node $s$} \Output{node ordering $\Delta$; set of filtered
	edges $E'$} Run Extended Dijkstra and take $\Delta$ and $E'$ \; \Return
	$\Delta, E'$
\end{algorithm}

{\bf \norat.} In constructing a DAG based on Dijkstra's algorithm, {\norad} only
takes into account the edges that generate shortest paths. On the other hand, in
{\norat}, method {\tt Analyze\-Network} transforms the original graph into a DAG
$G' = (V,E)$ where the edges in $E'$ do not necessarily have to be created from
edges that belong to shortest paths in $G$.
In this way, {\norat} attempts to consider more connections that may impact the companies. 

{\norat} performs a topological sorting and obtains a node ordering $\Delta$
(see Algorithm \ref{alg:filter-edges2}). If $G'$ has an edge $(n,m)$, the
topological ordering will place $n$ before $m$ \cite{cormen2009}. In line 2 of
Algorithm \ref{alg:nora}, the topological node ordering $\Delta$ ensures that a
node $n$ accumulates the values propagated to $n$ via all paths in $G'$.

The first step of {\norat} is to covert the input graph into a DAG. For
undirected graph, {\norat} starts by assigning a direction to each edge in the
graph, direction which corresponds to the one given by traversing the graph in a
breadth-first manner, starting with source node $s$. Any self-loop is also
detected and removed while traversing the graph, thus the result of this step is
a DAG. When the input graph is a directed graph, the initial step of {\norat} is
to remove the cycles in the graph. Eliminating a \emph{minimum} number of edges
from a directed cyclic graph to obtain a DAG is known as the feedback arc set
problem, which is NP-hard \cite{karp1972}. However, approximation techniques
that run in polynomial time exist. We take such an approximation approach, based
on Eades et al.'s algorithm~\cite{eades1993}. More specifically, {\norat}
computes a minimum feedback arc set using the Eades et al.'s
algorithm~\cite{eades1993} and then builds the final DAG $G'$ by removing from
$G$ all the edges in the set.

Either depth-first or breadth-first search can be used for the topological sort.
We employ depth-first search, as in \cite{tarjan1976,cormen2009}.

\begin{algorithm}[t!] \DontPrintSemicolon
	\small
	\caption{{\tt AnalyzeNetwork (\norat)}} \label{alg:filter-edges2}
	\SetKwInOut{Input}{input} \SetKwInOut{Output}{output} \Input{graph $G =
	(V,E)$, start node $s$} \Output{node ordering $\Delta$; set of filtered
	edges $E'$} Generate a DAG $G' = (V, E')$ from $G$ \; Perform topological
	sort and take $\Delta$ \; \Return $\Delta, E'$
\end{algorithm}


{\bf Worst-case Computational Complexity.} When using Dijkstra's algorithm, the
complexity of {\norad} is $O(|E| + |V| \log |V|)$, where $|V|$ is the number of
nodes and $|E|$ is the number of edges. Dijkstra's algorithm is the step with
the largest complexity.
%
When the graph edges have uniform costs, replacing Dijkstra's algorithm with
breadth-first search reduces {\norad}'s complexity to $O(|E| + |V|)$.
The complexity of topological sort and assigning direction to the edges is $O(|E| +
|V|)$. The algorithm of Eades \emph{et al.}~calculates a feedback arc set in
$O(|E|)$ time. {\norat} \cite{eades1993}, therefore, calculates flow values in
$O(|E| + |V|)$ time.
Recall that, after applying NORA to compute flows, nodes need to be ranked. The
ranking (sorting) complexity is within $O(|V| log(|V|))$. However, this step is
specific to the problem, rather than being specific to an algorithm such as
NORA. Regardless of the algorithm used to compute relevance scores for the
nodes, nodes would have to be ranked based on those scores.



\section{Evaluation Framework} \label{sec:evaluation}
Our evaluation goes into two main directions. Firstly, we evaluate
the algorithm by formulating it as a solution to link prediction. Secondly, we
gathered feedback from user interviews to evaluate the impact of our work from
the perspective of real users.

The Client Network can support sellers in identifying prospective clients by
highlighting the non-client companies which occupy the top positions in the
ranking computed by the ranking algorithm. A human expert is involved in
selecting the candidates, and contacting the selected candidates afterwards. A
human expert will not have time to process the entire list, and they typically
focus on a subset at the top of the list. Therefore, the \emph{Precision at K}
($P@K$) is a key metric for evaluating the performance of the ranking algorithm.
Specifically, we used Precision at 10, 50, 100 and 1000 for measuring the
quality of the recommendations.

Moreover, for the sake of completeness, we added to the comparison metrics that
are widely adopted in the link prediction
literature~\cite{DBLP:journals/corr/YangLC15}, such as the Area Under the
Receiver Operating Characteristic Curve (AUROC), the Area Under the
Precision-Recall Curve (AUPR) and the Top $|P|$ Predictive Rate
(TPR$_{|P|}$)\footnote{Where $P$ is the set of positive instances in the
prediction results~\cite{DBLP:journals/corr/YangLC15}.}.

The client relationship partitions the organization nodes into two distinct
classes, \emph{client} and \emph{non-client}. These two are unevenly represented
in the network. The imbalance ratio is 1 to 14 between client nodes and
non-client nodes. Furthermore, about 50\% of the client nodes are only connected
to the root node. This type of node does not provide any information that can be
leveraged by the link prediction algorithms, especially those that rely on the
topological structure of the network. Hence, only the remaining half of client
nodes together with the non-client ones are used in our experiments.

We compared NORA with two random-walk based methods, Rooted PageRank and PropFlow.
Rooted PageRank (RPR)~\cite{ASI:ASI20591} is a modified version of
PageRank. The rank of a node corresponds to the probability that the node
will be reached through a random walk from the source. A parameter $\alpha$
specifies how likely the algorithm is to visit the node's neighbors rather
than starting over.
PropFlow~\cite{Lichtenwalter:2010:NPM:1835804.1835837} is related to
Rooted PageRank, but it is more localized. The rank of a node corresponds to
the probability that a restricted random walk starting at the source node
ends at the target node in no more than $l$ steps. Unlike RPR, PropFlow does
not require walk restarts or convergence but simply employs a modified
breadth-first search restricted to depth $l$.

We used a 10-fold cross-validation stratified edge holdout scheme to measure the
performance of the three algorithms in predicting links of type client. A formal
description of the evaluation methodology is presented in
Algorithm~\ref{alg:framework}.
At line 1, method {\tt Sample\-Folds} randomly samples nodes from the client and
non-client classes. Its implementation ensures that the distribution of the two
classes is preserved in each fold.

Method {\tt Compute\-Metrics} (line 10 in Algorithm~\ref{alg:framework})
calculates the Precision at $10$, $50$, $100$, $1000$ as well as the AUROC, AUPR
and top $|P|$ predictive rate by labelling the client and non-client nodes in the
$i$th fold as positive and negative instances, respectively. These values are
computed for each algorithm $j$ and fold $i$ and stored into the
multidimensional vector {\tt Metrics}.

\begin{algorithm}
	\DontPrintSemicolon
	\small
	\caption{Evaluation Framework Workflow} \label{alg:framework}
	\KwData{graph $G = (V,E)$, root node $r$, set of link prediction algorithms $AS$}
	\KwResult{average values of AUROC, AUPR, top $|P|$ predictive rate and Precision at 10, 50, 100, 1000 for each algorithm in $AS$}
	$Folds \leftarrow \mbox{SampleFolds}(V)$ \;
	\For {each $fold_{i} \in Folds$} {
		$DC \leftarrow \emptyset$ \;
		\For {each $node \in fold_{i}$} {
			\If {node \textup{is a client}} {
				insert $node$ into $DC$ \;
				$\mbox{RemoveClientLink}(G, r, node)$ \;
			}
		}
		\For {each $alg_{j} \in AS$} {
			$Scores \leftarrow \mbox{ComputeScores}(G, r, alg_{j})$ \;
			$Metrics[j][i] \leftarrow \mbox{ComputeMetrics}(Scores, fold_{i})$ \; \label{pseudocode:compute-metrics}
		}
		\For {each $node \in DC$} {
			$\mbox{AddClientLink}(G, r, node)$ \;
		}
	}
	\For {each $row_{j} \in Metrics$} {
		$\mbox{ComputeAverageMetrics}(row_{j})$ \;
	}
\end{algorithm}

{\bf Results.}
All evaluated algorithms are implemented in Python using the NetworkX
module~\cite{Hagberg2008}. An \emph{undirected multigraph} 
represents the structure of the network. Further details on the configuration
parameters used in the experiments are available in Table~\ref{tab:params}.

Table~\ref{tab:metrics} contains the averages of the metrics for each algorithm.
NORA-D achieves the highest scores for the Precision at 10 and 50 while NORA-T
places the highest number of disconnected clients on the top 100 and 1000 of its
ranking. Rooted PageRank offers the highest performance on the AUROC, AUPR and
TPR$_{|P|}$ but it is also the worst in term of Precision when considering up to
position 1000 of the ranking. PropFlow never excels in any of the metrics
although it is never the worst.

PropFlow has a computational complexity of $O(|V| + |E|)$, where $|V|$ is the
number of nodes and $|E|$ the number of edges in the graph. So do both NORA
variants, since the graph has edges with uniform cost, as explained in
complexity analysis section.
Rooted PageRank is the algorithm
with the highest time complexity in the group since each $iteration$ of the
algorithm runs in linear time $O(|V| + |E|)$.

In summary, NORA has a good time complexity, and it performs well for the
Precision at K metrics. It is worth mentioning again that these metrics are very
important in our application since the list of companies that the Client Network
recommends the sellers to approach is extracted from the top of the ranking.

\begin{table}[h!]
	\small
	\centering
	\caption{Configurations used in the experiments}~\label{tab:params}
	\begin{tabular}{|l|l|}
		\hline
		Algorithm                        & Parameters                         \\
		\hline
		\hline
		\multirow{3}{*}{Rooted PageRank} & $\alpha$ = 0.15; Error tolerance = $1.0e-6$                    \\
		                                 & Maximum number of iterations = 100 \\
		\hline
		PropFlow                         & Depth = 10                         \\
		\hline
		NORA (both variants)             & Discount factor = 0.95             \\
		\hline
		{All}             & Edge cost = 1.0; Edge weight\footnotemark = 1.0                    \\
		\hline
	\end{tabular}
\end{table}
\footnotetext{We use the term ``weight'' for indicating the flow capacity associated to each edge.}

\begin{table*}[t]
	\centering
	\caption{Client Prediction Results\protect\footnotemark}~\label{tab:metrics}
	\begin{tabular}{|l|c|c|c|c|c|c|c|}
		\hline
		Algorithm       & $P@10$        & $P@50$         & $P@100$        & $P@1000$       & TPR$_{|P|}$    & AUROC          & AUPR           \\
		\hline\hline
		Rooted PageRank & 0.39          & 0.55           & 0.578          & 0.589          & \textbf{0.338} & \textbf{0.917} & \textbf{0.292} \\
		\hline
		PropFlow        & 0.81          & 0.73           & 0.715          & 0.628          & 0.313          & 0.833          & 0.26           \\
		\hline
		NORA-D          & \textbf{0.85} & \textbf{0.746} & 0.722          & 0.617          & 0.265          & 0.816          & 0.214          \\
		\hline
		NORA-T          & 0.82          & 0.734          & \textbf{0.728} & \textbf{0.634} & 0.284          & 0.823          & 0.232          \\
		\hline
	\end{tabular}
\end{table*}

\footnotetext{The values are averages over the 10 folds. In bold the highest value for each metric.}
\section{User interviews} \label{sec:user-interviews}
Our aim in these interviews was to understand which features of the application
were considered the most and least useful by the users and which features could
achieve a time saving when approaching prospective clients. We aimed to
understand if the visualization and the ranking would aid the user in their
daily tasks.

{\bf Method.} We demonstrated the system in the sales department in our
organization, after which we recruited 5 volunteers whose job roles involve
evaluating whitespace clients. They were asked to evaluate the usability and
usefulness of the Client Network in relation to their daily tasks. After initial
group demonstrations, they tried out the functionality. We followed up with
semi-structured interviews. A questionnaire with open-ended questions was
developed which allowed some scope for exploring some of the responses in depth.
We found this approach useful as the sellers have different roles in the
organization and therefore were using the Client Network in different ways.

{\bf Results.} The sellers were enthusiastic about the functionality of the
Client Network, but they also pointed out a need for a deeper integration with
the Sales Customer Relationship Management (CRM) tool they use before they can
really make use of its time-saving potential. There were positive responses to
the company search coverage, indicating that they could find results related to
the companies they were searching for, but there were some omissions when they
searched for public sector clients.

Sellers showed an interest in the application's display of information about the
movement of people and subsidiaries. This information is difficult, if not
impossible, to uncover in their existing software toolset, and is useful for
them to know before they contact the company. Several sellers thought that the
company description is very useful information to display because
\textit{``\dots it's important to know a bit about the client when you are
calling the client\dots''}, and that this information could be useful as a
conversational ``hook'' when cold calling \textit{``\dots when you call 20
people a day you obviously cannot research every individual in a very detailed
way.''}

The majority of sellers reported that the visualization was the most important
part of the application to them from an exploratory point of view. A number of
key requirements have been identified for successful network exploration tools,
including high quality layout algorithms, data filtering, clustering, statistics
and annotation~\cite{bastian2009gephi}. The findings of the user interviews
suggest that we should reconsider our approach to annotating the data, and that
showing the visualizations as the only interface to the data should be
reconsidered if we are considering a task-based rather than exploratory use, to
afford users the option of exploring visually or in a more structured manner.

There were very positive comments about the time-saving aspect of showing the
company and subsidiary information, and the fact that the data was more up to
date than the existing system that the seller would look in for this type of
information. UI improvements that were suggested were to add a legend to the
visualization to help interpret the different type of connections and to improve
the graph search filters.

There was a strong belief from the majority of the sellers that the prototype
could be even more useful if it was expanded to display historical product
information when showing information about linked companies, for example, if
there was a linkage between nodes on the graph and the sales CRM that is used in
the company so that you could see which products the connected clients had
purchased previously \textit{``\dots there is a lot of functionality that if
they are expanded in the right way, can actually help us a lot, and integrations
are the very zero point to start\dots''}. Another case which was suggested for
the graph is to highlight the companies that have business relationships, as
this is helpful when the sellers are trying to target business-to-business
products.
\section{Conclusion and Future work} \label{sec:conclusion}
In this paper, we formalized the problem of identifying prospective clients in
sales and proposed an innovative solution, Client Network. It utilizes a novel
ranking algorithm for predicting new clients and provides an interactive
interface for allowing the exploration of the client ecosystem. While typical
approaches study client interactions or explore clients' similarity to existing
clients based on market segmentation, our approach leverages the whole structure
of the client ecosystem. This ecosystem can be represented as a heterogeneous
network and, in such a context, the problem of identifying prospective clients
can be formalized as an instance of the link prediction problem. We presented
NORA, a novel ranking algorithm and compared its performance with two well known
algorithms in the link prediction field. The experiments demonstrate that our
technique achieves higher precision than Rooted PageRank and PropFlow. Our
approach can be considered highly suitable for the use cases covered by the
Client Network. In user interviews sellers expressed their appreciation about
how the Client Network can help them in prioritizing leads and how the
information it provides is complementary to that available from the tools they
use to work with. As part of future work, we will iterate the design of the UI
based on the user feedback, add new features based on new use cases and explore
more novel ways to display the subgraph. Furthermore, we aim to further improve
the quality of the recommendations and re-evaluate the performance of the
algorithms with new experiments. 
\section{Acknowledgements}
We would like to acknowledge the support and collaboration of the CAO team:
Alice Chang, Claire Tian, Ben J Dubiel, Sanjmeet Abrol and Weiwei Li; for their
valuable insights into the realm of digital sellers.

%
\bibliographystyle{splncs04}
\bibliography{references}

\end{document}